\documentclass[aps,pre,reprint]{revtex4-1}
\usepackage{blindtext}
\usepackage{graphicx}
\usepackage{dcolumn}
\usepackage{bm}
\usepackage{xcolor}
\usepackage{amsmath}
\usepackage{amssymb}
\usepackage{appendix}
\usepackage[urlcolor=blue]{hyperref}

\begin{document}
\preprint{APS/123-QED}

\title{Flocking through a sea of rods}
\author{Abhishek Sharma}
\author{Harsh Soni}
\email{harsh@iitmandi.ac.in}
\affiliation{School of Physical Sciences (SPS), Indian Institute of Technology Mandi, Parashar Road, Tehsil Sadar, Near Kataula, Kamand, Himachal Pradesh 175005, India.}

\begin{abstract}
We investigate the collective behavior of motile rods immersed in a monolayer of apolar rods confined between vertically vibrating plates using numerical simulations. We uncover an antidiffusive instability whereby motile rods segregate from the apolar medium and form flocks whose size increases with the medium concentration. Remarkably, enhanced segregation leads to a reduction of the global polar order. The flock structure is strongly influenced by the anisotropy of the medium rods. For small aspect ratios, the flocks are elongated perpendicular to the mean direction of motion, whereas for larger aspect ratios they elongate along the direction of motility. We rationalize the emergence of segregation-induced disorder using a minimal mean-field model.
\end{abstract}

\maketitle
Introduction: Flocking is a central phenomenon of active matter, emerging when motile entities interact through a surrounding medium, direct mechanical contact, or perceptual cues, giving rise to robust collective motion and striking spatiotemporal patterns~\cite{toner2005hydrodynamics,ramaswamy2010mechanics,RevModPhys.85.1143,paoluzzi2024flocking,toner2024physics,giraldo2025active,luo2025flocking,fava2024strong,yang2010swarm,nagai2018collective,kuan2015hysteresis,kursten2025emergent}. It displays intrinsically nonequilibrium features absent in passive systems and occurs across an extraordinary range of physical realizations, from living organisms spanning multiple length scales to synthetic motile particles engineered in the laboratory. In this Letter, we investigate how a medium composed of elongated particles influences flocking in dry active granular systems using numerical simulations. While most studies focus on flocking in structureless environments~\cite{VICSEK201271,Schaller2010,doi:10.1073/pnas.1001651107,PhysRevLett.108.098102,PhysRevLett.102.010602,PhysRevE.74.030904,PhysRevLett.105.098001,RevModPhys.85.1143,PhysRevLett.75.1226,PhysRevLett.92.025702,ramaswamy2003active,kumar2014flocking,4}, far less is known about how flocking emerges when the surrounding medium is complex~\cite{PhysRevLett.110.238101,PhysRevLett.111.160604,PhysRevE.106.064602,rahmani2021topological,PhysRevE.110.L062102,chepizhko2015active,aceves2020large,sampat2021polar,vahabli2023emergence,PhysRevE.105.064612,mondal2025dynamical,zhou2017dynamic,rajabi2021directional,boule2020dynamic,crj5-bhwv,goral2022frustrated,nyvr-knp8,bechinger2016active,lopes2025emergence,PhysRevX.14.031008,aranson2018harnessing,sokolov2015individual,peng2016command,zhou2014living,PhysRevX.7.011029,dadwal2025nematic,kumar2022catapulting,dadwal2023quantifying}. Motivated by recent studies of polar swimmers navigating nematic backgrounds in both wet and dry active-matter systems~\cite{sampat2021polar,mondal2025dynamical,zhou2017dynamic,rajabi2021directional,boule2020dynamic,crj5-bhwv,goral2022frustrated,nyvr-knp8,sokolov2015individual}, we address how a collection of granular, motile (polar)
rods flocks while traversing a monolayer medium of granular, nonmotile (apolar) rods confined between two vertically shaken plates with periodic boundary conditions in the horizontal plane. This medium exhibits a rich variety of phases--including isotropic, nematic, chevron, and defect-melted nematic states--accessible by tuning the rod length and concentration~\cite{pahsesAS_HS}. Such tunability introduces an additional layer of complexity to the emergent dynamics, significantly enriching the system’s behavior.


We summarize our main findings as follows.
We find that motile rods exhibit a strong tendency to segregate from the apolar medium and form flocks as the medium concentration increases. Enhanced segregation suppresses the global polar order of the motile rods. In the absence of intrinsic angular noise, this mechanism leads to a phase transition from a homogeneous polar-ordered state to a disordered, segregated regime characterized by elongated flocks that do not sustain global order. 
Introducing intrinsic angular noise progressively weakens segregation and, counterintuitively, enhances polar order, revealing an anomalous noise-induced ordering effect. At sufficiently large noise strengths, the system ultimately crosses over to a homogeneous disordered phase.
At high apolar-rod concentrations, the flock structure is strongly influenced by the anisotropy of the medium rods. For small aspect ratios, the flocks are elongated perpendicular to the mean direction of motion, whereas for larger aspect ratios they align parallel to the direction of motility, consistent with segregation-induced structures observed previously in a bead medium~\cite{kumar2014flocking}. We rationalize the emergence of segregation-induced disorder using a minimal mean-field model.


Methods: We perform Newtonian dynamics simulations of our system of rigid particles confined between vertically oscillating plates, in which both interparticle and particle--wall collisions are assumed to be inelastic~\cite{kumar2014flocking,2,3,4,5,pahsesAS_HS,motilerod_AS,dabra2025depletion,narayan2007long}. The vertically oscillating plates are modelled as horizontally extended moving walls with vertical positions $A\cos(\Omega t)$ and $A\cos(\Omega t)+w$, respectively, where $A$ denotes the vibration amplitude, $\Omega$ the angular frequency, and $w=1.2\,\mathrm{mm}$ is the fixed gap between the plates. The simulations are further simplified by imposing periodic boundary conditions in the $xy$ plane. Collisions are treated using an impulse-based collision framework~\cite{Stronge1994}. Each rod is represented as a linear chain of partially overlapping spheres.The polar rods have a fixed length \(\ell_{\mathrm{p}} = 4.5\,\mathrm{mm}\) and a diameter that varies from \(1.375\,\mathrm{mm}\) at the thick end to \(0.4\,\mathrm{mm}\) at the thin end. The apolar rods are symmetrically tapered at both ends, with a diameter varying from \(0.8\,\mathrm{mm}\) at the center to \(0.4\,\mathrm{mm}\) at the ends, and have lengths \(\ell_{\mathrm{a}}\) ranging from \(0.8\,\mathrm{mm}\) to \(4.5\,\mathrm{mm}\), corresponding to aspect ratios \(\sigma\) varying from \(1\) to \(5.625\). Note that \(\ell_{\mathrm{a}} = 0.8\,\mathrm{mm}\) corresponds to a spherical bead.
 The friction and restitution coefficients are chosen to be the same as the values established in previous studies of polar rods in a bead medium~\cite{4}. The dimensionless acceleration parameter $\Gamma=A\Omega^{2}/g$ is set to $7$, with the driving frequency fixed at $\Omega=400\pi\,\mathrm{s^{-1}}$, where $g$ is the gravitational acceleration. We also incorporate rotational noise for the polar rods by introducing a random angular velocity in the $xy$ plane upon collisions with the plates, quantified in terms of the single-rod rotational diffusion coefficient $D_{\mathrm{r}}$.
The simulation box size is set to $L=56\,\ell_{\mathrm{p}}$, unless stated otherwise. The concentrations of polar and apolar rods are measured by their respective area fractions, $\phi_{\mathrm{p}}$ and $\phi_{\mathrm{a}}$, defined as the fraction of the system area covered by the rods, assuming they lie entirely in the $xy$ plane. Visualization and rendering of simulation trajectories are performed using the VMD software package~\cite{VMD}. Various averaged quantities calculated here are estimated using time averages in the steady state. Additional details---including friction and restitution coefficients, rod geometries, rod numbers, the precise definition of area fraction, and the mechanism of rotational noise---are provided in the Supplemental Material (Sec.~\ref{sec:sim_details})~\cite{motilerod_AS}.

\begin{figure}[h!]
    \centering
    \includegraphics[width=1.0\linewidth]{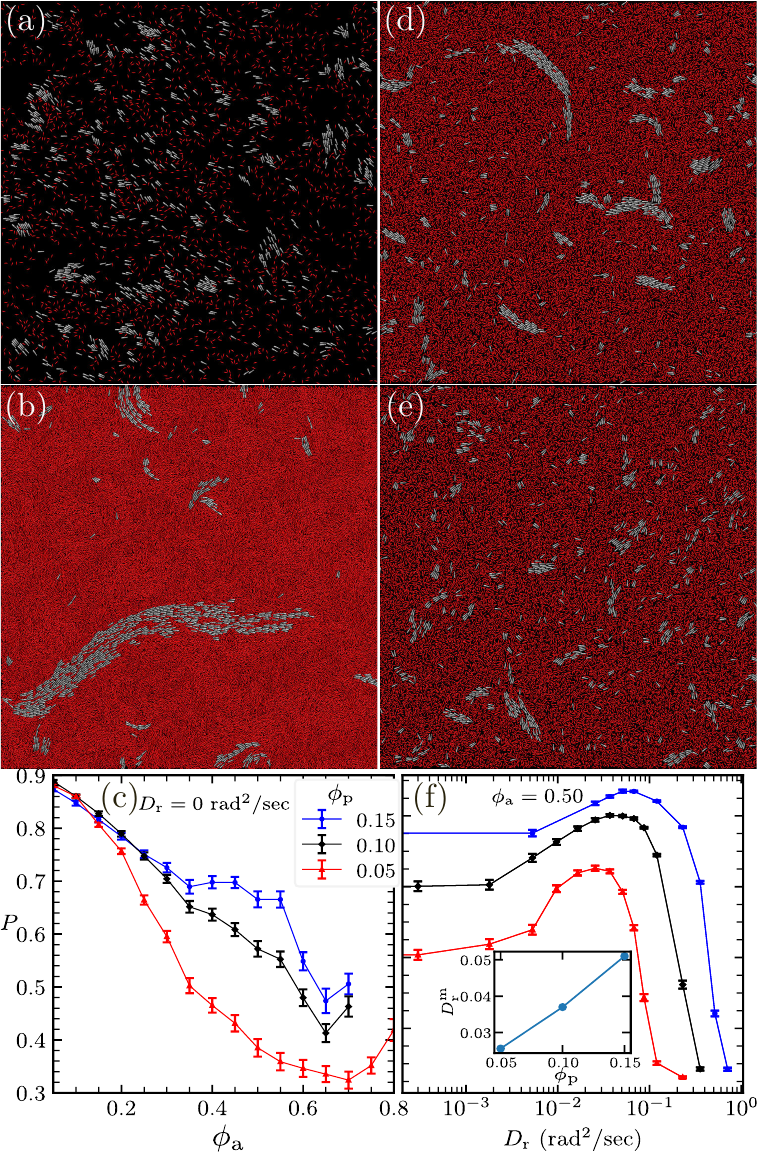}
   \caption{\textbf{Effect of the area fraction \(\phi_{\mathrm{a}}\) of apolar rods at \(D_{\mathrm{r}} = 0~\mathrm{rad}^2\,\mathrm{s}^{-1}\).} (a,b) Steady-state configurations at \(\phi_{\mathrm{p}} = 0.05\) for (a) \(\phi_{\mathrm{a}} = 0.05\) and (b) \(\phi_{\mathrm{a}} = 0.80\). (c) Polar order parameter \(P\) of the motile rods vs \(\phi_{\mathrm{a}}\) for different \(\phi_{\mathrm{p}}\). \textbf{Effect of the intrinsic angular noise \(D_{\mathrm{r}}\) of the polar rods at \(\phi_{\mathrm{a}} = 0.50\).} (d,e) Steady-state configurations at \(\phi_{\mathrm{p}} = 0.05\) for (d) \(D_{\mathrm{r}} = 0~\mathrm{rad}^2\,\mathrm{s}^{-1}\) and (e) \(D_{\mathrm{r}} = 0.026~\mathrm{rad}^2\,\mathrm{s}^{-1}\) (corresponding to the maximum value of \(P\) in (f)). (f)  \(P\) vs \(D_{\mathrm{r}}\) for different \(\phi_{\mathrm{p}}\). Inset: $D_{\mathrm{r}}^{\mathrm{m}}$ (where $P$ is maximal) vs $\phi_{\mathrm{p}}$.  Here $\sigma=0.3125$ (\(\ell_{\mathrm{a}} = 2.5\,\mathrm{mm}\)). The apolar-rod medium remains in the isotropic phase.  
} 
    \label{fig01}
\end{figure}
Results: We first investigate the collective dynamics of motile rods embedded in an isotropic medium at $\sigma=3.12$ (\(\ell_{\mathrm{a}} = 2.5\,\mathrm{mm}\)) in the absence of intrinsic angular noise \(\left(D_{\mathrm{r}} = 0~\mathrm{rad}^2\,\mathrm{s}^{-1}\right)\). The plots of the nematic order parameter for the medium, shown in the Supplementary Material [Sec.~\ref{sec:nematic_odpm}, Fig.~\ref{figS}(a)], confirm that the medium is in an isotropic phase.
Fig.~\ref{fig01}(a) shows that at low \(\phi_{\mathrm{a}}\), the motile rods form small flocks that move coherently, resulting in an ordered phase. In contrast, at large \(\phi_{\mathrm{a}}\), larger flocks emerge that move incoherently, leading to a disordered phase [Fig.~\ref{fig01}(b), Supplementary Movie SM1]. These flocks are highly dynamic, continuously breaking into smaller flocks that subsequently merge to form larger polar flocks (see Supplemental Material, Sec.~\ref{sec:clusterana}, Fig.~\ref{figlg}). The instantaneous direction of flock motion is strongly influenced by emergent topological defects in the medium (see Supplementary Movie SM1), and the flocks develop elongated tails along their direction of motion, a structure we term \emph{longitudinal flocking}.  A detailed analysis of the flock size and shape is presented later. The transition from an ordered to a disordered phase with increasing \(\phi_{\mathrm{a}}\) is further confirmed in Fig.~\ref{fig01}(c), which shows that the polar order parameter \(P\) decreases as \(\phi_{\mathrm{a}}\) increases for different values of \(\phi_{\mathrm{p}}\).
Strikingly, this trend contrasts with that observed in a spherical-bead medium, where the medium instead enhances ordering at large $\phi_{\mathrm{a}}$ (see Supplemental Material, Sec.~\ref{sec:polar_sphericalbeads}, Fig.~\ref{figbead})~\cite{kumar2014flocking,4}.  Moreover, rods in a bead medium form a homogeneous ordered phase, in contrast to the randomly moving clustered flocks observed here.

\begin{figure}[!h]
    \centering
    \includegraphics[width=\linewidth]{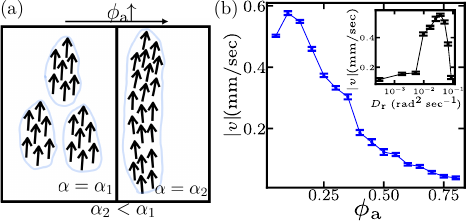}
    \caption{(a) Schematic illustrating the mechanism of segregation-induced disorder. As segregation between motile and apolar rods increases with $\phi_{\mathrm{a}}$, the interfacial length decreases, leading to a reduction of the coupling constant $\alpha$ in Eq.~\eqref{eq:v_eom}. 
(b) Average speed $|\mathbf{v}|$ of the medium rods measured at a distance of $100\,\mathrm{mm}$ from a polar rod in its comoving frame, averaged over rods and time, as a function of $\phi_{\mathrm{a}}$ for $D_{\mathrm{r}} = 0~\mathrm{rad}^2\,\mathrm{s}^{-1}$. Consistent with the mechanism shown in (a), $|\mathbf{v}|$ decreases with increasing $\phi_{\mathrm{a}}$. Inset: $|\mathbf{v}|$ as a function of $D_{\mathrm{r}}$ at $\phi_{\mathrm{a}} = 0.50$, showing an increase at small $D_{\mathrm{r}}$.
Here, $\sigma = 3.125$ ($\ell_{\mathrm{a}} = 2.5\,\mathrm{mm}$) and $\phi_{\mathrm{p}} = 0.05$.
}
    \label{fig05}
\end{figure}

Two mechanisms underlie this anomalous medium-induced behavior. First, as \(\phi_{\mathrm{a}}\) increases, the coupling between the orientations of the motile rods and the local nematic alignment of the medium enhances the effective rotational diffusion of the rods. This effect becomes increasingly pronounced at larger \(\phi_{\mathrm{a}}\), where topological defects in the medium play a significant role [see Supplemental Material, Sec.~\ref{sec:nematic_odpm}, Fig~\ref{figCtC2t}(c)]. Second, strong medium-mediated attraction between the motile rods promotes their coalescence into larger flocks, thereby reducing the interfacial length between the polar-rod-rich regions and the surrounding medium (Fig.~\ref{fig05}a). The reduced interfacial length weakens the ability of the motile rods to push the medium, suppressing the medium-induced velocity field responsible for enhanced ordering [see Fig.~\ref{fig05}(b)]~\cite{kumar2014flocking}. A simple theoretical model based on the mechanisms discussed above is presented in the End Matter.

\begin{figure}[!h]
    \centering
     \includegraphics[width=1.0\linewidth]{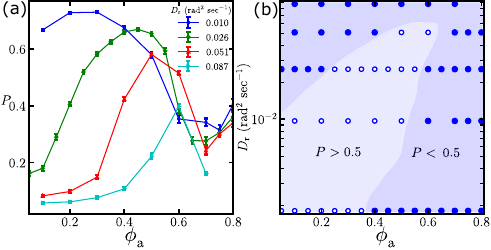}
    \caption{ (a) $P$ vs $\phi_{\mathrm{a}}$ for different $D_{\mathrm{r}}$. 
(b) Phase diagram in the $D_{\mathrm{r}}$--$\phi_{\mathrm{a}}$ plane. Hollow and solid circles denote states with $P < 0.5$ and $P > 0.5$, respectively. 
Here, $\sigma = 0.3125$ ($\ell_{\mathrm{a}} = 2.5\,\mathrm{mm}$) and $\phi_{\mathrm{p}} = 0.05$.
}
    \label{fig02}
\end{figure}

Next, we examine the role of intrinsic angular noise \(D_{\mathrm{r}}\) of the polar rods, keeping \(\phi_{\mathrm{a}} = 0.50\) fixed, in the same isotropic medium ($\sigma=3.12$). As discussed above, at \(D_{\mathrm{r}} = 0~\mathrm{rad}^2\,\mathrm{s}^{-1}\), the system is in a disordered phase with large, uncorrelated flocks [see Fig.~\ref{fig01}(d), Supplementary Movie SM2]. Upon introducing a small amount of angular noise, the large flocks fragment into smaller ones, increasing the interfacial length between the polar-rod-rich regions and the surrounding medium. The increased interfacial length enables the polar rods to drive the medium more effectively, thereby enhancing the medium velocity field [inset of Fig.~\ref{fig05}(b)] and increasing polar ordering [see Fig.~\ref{fig01}(e) and Supplementary Movie SM2]. At larger values of \(D_{\mathrm{r}}\), however, angular noise dominates over alignment induced by the velocity field, causing the polar order to decrease and leading to a reentrant disordered phase. Fig.~\ref{fig01}(f) shows that \(P\) initially increases with  \(D_{\mathrm{r}}\) and then decreases beyond an optimal value \(D_{\mathrm{r}}^\mathrm{m}\).
This nonmonotonic trend persists across the range of \(\phi_{\mathrm{p}}\) studied, and \(D_{\mathrm{r}}^\mathrm{m}\) increases with \(\phi_{\mathrm{p}}\) [see the inset of Fig.~\ref{fig01}(f)].
For completeness, we plot $P$ as a function of $\phi_{\mathrm{a}}$ in Fig.~\ref{fig02}(a) for different $D_{\mathrm{r}}$ and observe a similar nonmonotonic dependence at finite $D_{\mathrm{r}}$, consistent with the same underlying mechanism.
 Fig.~\ref{fig02}(b) shows the phase diagram in the \(D_{\mathrm{r}}\)-\(\phi_{\mathrm{a}}\) plane, where states with \(P < 0.5\) are denoted by solid blue circles and those with \(P > 0.5\) by hollow blue circles. 

We now study the shape and size of the flocks and the degree of segregation in the system. We compute the segregation order parameter \(\Sigma \in [0,1]\), where \(\Sigma = 0\) corresponds to the theoretically minimal level of segregation, while \(\Sigma = 1\) indicates complete segregation~\cite{mccandlish2012spontaneous,bhowmik2025segregation} (see Supplemental Material, Sec.~\ref{sec:SegODPM}).
\begin{figure}[!h]
    \centering
     \includegraphics[width=1.0\linewidth]{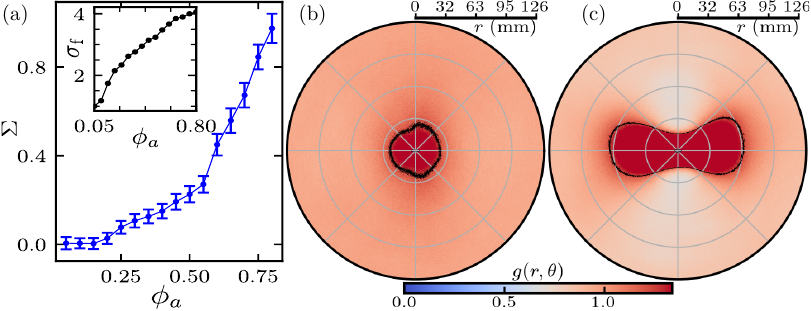}
    \caption{(a) Segregation order parameter $\Sigma$ vs $\phi_{\mathrm{a}}$; the inset shows $\sigma_{\mathrm{f}}$, quantifying the flock aspect ratio, as a function of $\phi_{\mathrm{a}}$. 
Polar heat maps of $g(r,\theta)$ at (b) $\phi_{\mathrm{a}} = 0.05$ and (c) $\phi_{\mathrm{a}} = 0.80$. 
Here, $\sigma = 0.3125$ ($\ell_{\mathrm{a}} = 2.5\,\mathrm{mm}$), $D_{\mathrm{r}} = 0~\mathrm{rad}^2\,\mathrm{s}^{-1}$, and $\phi_{\mathrm{p}} = 0.05$.
}
    \label{fig03}
\end{figure}
Fig.~\ref{fig03}(a) shows that \(\Sigma\) increases with \(\phi_{\mathrm{a}}\) at $D_r=0~\mathrm{rad}^2\,\mathrm{s}^{-1}$, indicating that the system becomes increasingly segregated into polar-rod-rich and apolar-rod-rich regions at larger \(\phi_{\mathrm{a}}\). This observation is consistent with our earlier finding that the flock size increases with \(\phi_{\mathrm{a}}\) [see Fig.~\ref{fig01}(a,b)] and arises from the strong effective attraction between polar rods mediated by the apolar medium. A detailed analysis of this mechanism will be presented elsewhere~\cite{AS_HS_preprint}.
To examine the shape of the flocks, we compute the pair distribution function \(g(r,\theta)\) in the corotating frame of a reference polar rod, averaged over all polar rods. Figs.~\ref{fig03}(b) and (c) show polar heat maps of \(g(r,\theta)\) for $\sigma=3.12$ and \(\phi_{\mathrm{p}} = 0.05\) at \(\phi_{\mathrm{a}} = 0.05\) and \(0.80\), respectively. At low \(\phi_{\mathrm{a}}\), the distribution \(g(r,\theta)\) is isotropic, indicating that the flocks are somewhat isotropic. In contrast, at large \(\phi_{\mathrm{a}}\), the flocks become longitudinal, with \(g(r,\theta)\) strongly anisotropic and exhibiting an enhanced probability of finding polar rods ahead of and behind the reference rod over large distances, resulting in elongated structures.
We further assess this elongation by computing the average aspect ratio of the flocks $\sigma_{\mathrm{f}}$, obtained from the closed contour corresponding to \(70\%\) of the maximum value of \(g(r,\theta)\) [black contours in Figs.~\ref{fig03}(b) and (c)]; as shown in the inset of Fig.~\ref{fig03}(a), $\sigma_{\mathrm{f}}$ increases with $\phi_{\mathrm{a}}$. Orientational correlations, quantified by the polar order--parameter correlation function \(G(r,\theta)\), display a similar anisotropic structure at large $\phi_{\mathrm{a}}$, with strong alignment along the longitudinal direction and weaker correlations transverse to it (see Supplemental Material, Sec.~\ref{sec:nematic_odpm}, Fig.~\ref{figGrtheta}).

\begin{figure}[!h]
    \centering
    \includegraphics[width=\linewidth]{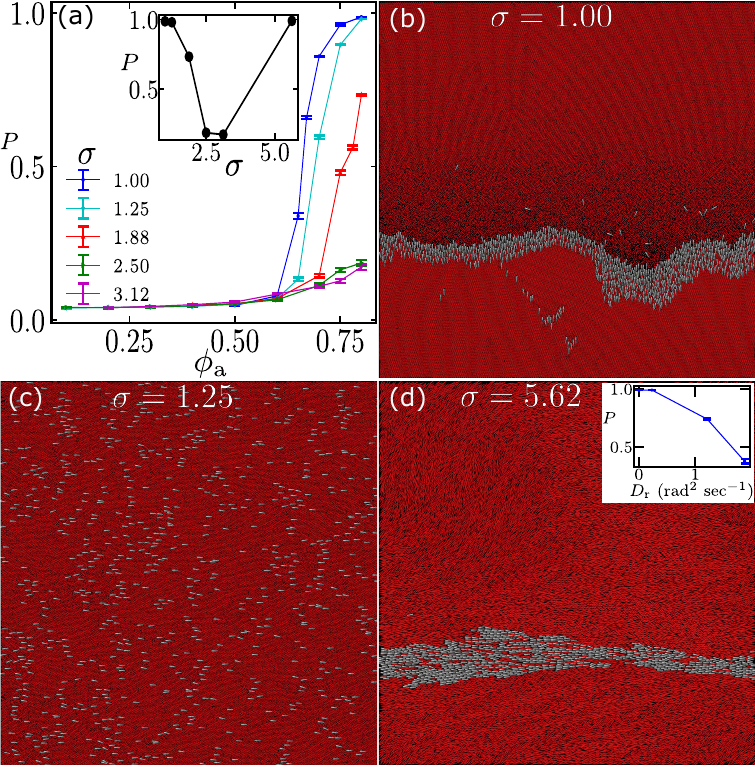}
    \caption{(a) $P$ vs $\phi_{\mathrm{a}}$ for different values of $\sigma$. Also, see Fig.~\ref{fig06} illustrating the phase diagram in the $\sigma-\phi_\mathrm{a}$ plane.
(b)--(d) Steady-state configurations for (b) $\sigma = 1.0$, (c) $\sigma = 1.25$, and (d) $\sigma = 5.62$. 
Here, $D_{\mathrm{r}} = 0~\mathrm{rad}^2\,\mathrm{s}^{-1}$ and $\phi_{\mathrm{p}} = 0.05$. 
Inset of (d): $P$ vs $D_{\mathrm{r}}$ for $\phi_{\mathrm{p}} = 0.05$.}
    \label{fig04}
\end{figure}

\begin{figure}[!h]
    \centering
     \includegraphics[width=0.9\linewidth]{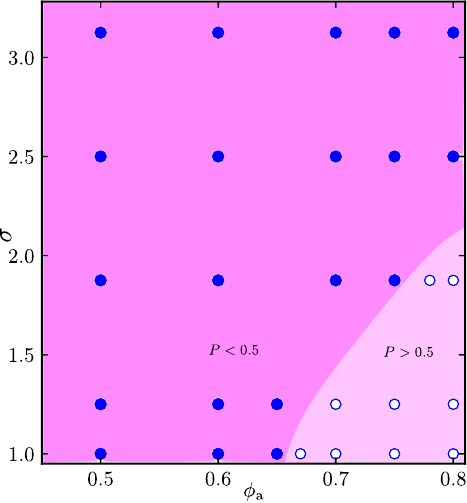}
    \caption{Phase diagram in the $\sigma$--$\phi_{\mathrm{a}}$ plane corresponding to Fig.~\ref{fig04}(a).}
    \label{fig06}
\end{figure}

To elucidate the role of medium-rod anisotropy in the order-disorder phase transition, we set $D_{\mathrm{r}} = 0.23~\mathrm{rad}^2\,\mathrm{s}^{-1}$, for which polar rods are disordered in the absence of the medium. We then vary $\sigma$ and  $\phi_{\mathrm{a}}$, while keeping $\phi_{\mathrm{p}} = 0.05$ fixed. Fig.~\ref{fig04}(a) presents $P$ vs $\phi_{\mathrm{a}}$  for different $\sigma$. For small $\sigma$, $P$ increases with $\phi_{\mathrm{a}}$, indicating a transition from a disordered to an ordered phase, consistent with earlier observations for polar rods in a bead medium ($\sigma = 1$)~\cite{kumar2014flocking}.
In contrast, increasing $\sigma$ within the isotropic phase of the apolar medium suppresses polar ordering at large $\phi_{\mathrm{a}}$, thereby inhibiting the phase transition. As displayed in the inset of Fig.~\ref{fig04}(a), at $\phi_{\mathrm{a}} = 0.75$, $P$ decreases with $\sigma$ up to $\sigma = 3.125$, for which the medium remains isotropic. Upon increasing $\sigma$ further to 5.625, the apolar medium becomes ordered, and the polar rods recover orientational order.
Interestingly, at large $\phi_{\mathrm{a}}$, the structure of the flocks depends sensitively on $\sigma$. For $\sigma = 1$ (the bead-medium case), the polar rods form transverse band, whose extent perpendicular to the mean direction of motion exceeds that along it, consistent with~\cite{4} [see Fig.~\ref{fig04}(b) and Supplementary Movie~SM3].
Increasing \(\sigma\) to \(1.25\) results in a homogeneous spatial distribution despite polar order [see Fig.~\ref{fig04}(c) and Supplementary Movie~SM4]. For intermediate values of \(\sigma\), global order is destroyed, and the system exhibits small, randomly moving longitudinal flocks [Supplementary Movie~SM4]. At sufficiently large \(\sigma\), where the medium itself has nematic ordering, polar order is restored, and the polar rods form a long band oriented along the direction of motion, extending across the simulation box [see Fig.~\ref{fig04}(d) and Supplementary Movie~SM4]. 
A phase diagram in the $\sigma$--$\phi_{\mathrm{a}}$ plane is displayed in Fig.~\ref{fig06}.
The band develops undulations, leading to the creation of topological defects in the medium that, in turn, drive the band through active nematic currents. As a result, the band begins to drift in a direction normal to its orientation (Supplementary Movie~SM5). At sufficiently large $D_{\mathrm{r}}$, the band progressively dilutes, and at still larger $D_{\mathrm{r}}$ the system becomes disordered, accompanied by a loss of nematic order in the medium [see inset of Fig.~\ref{fig04}(d) and  of Supplementary Material, Sec.~\ref{sec:innematicphase}, Fig.~\ref{figl45n15n30n36}].

Discussion: Defect dynamics plays a central role in the behavior observed in this system. The dynamics of elongated flocks is strongly influenced by the presence of topological defects in the medium. In particular, in the nematic medium, defect creation and annihilation events lead to the fragmentation and breakup of bands, resulting in strong temporal fluctuations of the polar order parameter [Supplementary Movie SM4].
A fundamental open problem is to understand the role of nematic defects in this behavior and to study the interfacial instability at the boundary between polar-rich and apolar-rich regions that may drive the breakup of these bands. The flocks also exhibit rich collective dynamics, including coherent reorientation events such as smooth global turns reminiscent of biological flocks [Supplementary Movie SM4], as well as occasional milling-like motion [Supplementary Movie SM5]. Remarkably, the dynamics of each flock is often governed by a leading polar rod at the front of the flock, which acts as an effective leader. Reorientation initiated by this leading rod propagates coherently through the entire flock, producing a collective turn of the whole structure. At high \(\phi_{\mathrm{p}}\), such collective turning events are accompanied by the development of density gradients in the apolar medium.
In the nematic medium,  at large \(\phi_{\mathrm{a}}\), we further observe the periodic formation of bands associated with a layered phase of the medium~\cite{pahsesAS_HS}.



In summary, our simulations demonstrate that motile rods immersed in an  medium of apolar rods undergo segregation-driven collective behavior. Increasing the medium concentration promotes the formation of segregated flocks while simultaneously diminishing global polar order. In the absence of intrinsic angular noise, this interplay gives rise to a transition from a homogeneous polar-ordered state to a disordered regime composed of elongated, segregated flocks lacking system-spanning order. 
Remarkably, angular noise counteracts segregation and enhances polar order over an intermediate range, leading to an anomalous noise-induced ordering effect before the system ultimately becomes homogeneously disordered at large noise strengths. These findings are captured qualitatively by a minimal mean-field framework, which rationalizes the emergence of segregation-induced disorder.
We further show that the geometry of the flocks is governed by the anisotropy of the medium rods: small aspect ratios favor structures elongated transverse to the mean direction of motion, whereas larger aspect ratios produce flocks aligned with the direction of motility~\cite{kumar2014flocking}. Thus, the medium plays a crucial role in shaping the anisotropic character of the antidiffusive behavior. 

AS acknowledges the computational resources provided by Param Himalaya under the National Supercomputing Mission (NSM) and financial support from the Prime Minister's Research Fellows (PMRF) scheme. HS acknowledges financial support from SERB through the SRG grant (No. SRG/2022/000061-G).

\bibliographystyle{apsrev4-1} 
\bibliography{main} 
\clearpage
\onecolumngrid
\section*{End Matter}

\twocolumngrid

Theory:  In our theoretical description, we do not address the antidiffusive mechanism responsible for the formation of segregated flocks, as this would require a detailed continuum theory involving coupled equations for the polar order parameter field, the nematic order parameter field, the velocity field, and the polar and apolar area-fraction fields. Such a framework, previously developed for bead media~\cite{kant2024bulk}, would need to be extended to incorporate the nematic order parameter in order to account for the formation of longitudinal bands or flocks. While a complete theory lies beyond the scope of the present work, the essential physical origin can be traced to the effective attraction between motile polar rods mediated by the apolar medium.
Here, we focus solely on the order--disorder phase transition
and restrict ourselves to the empirical observation that polar and apolar rods exhibit a strong tendency to segregate. We therefore construct a minimal mean-field description involving only the spatially averaged polar order parameter $\mathbf{P}$ and the mean velocity $\mathbf{v}$, following Kumar \textit{et al.}~\cite{kumar2014flocking}. The resulting equations of motion are given by
\begin{align}
\partial_t \mathbf{P} &= -a\,\mathbf{P} + \lambda\,\mathbf{v}
- b \lvert \mathbf{P} \rvert^2 \mathbf{P}, \label{eq:P_eom} \\
\rho\,\partial_t \mathbf{v} &= -\Gamma\,\mathbf{v} + \alpha\,\mathbf{P},
\label{eq:v_eom}
\end{align}
Here, $\rho$ denotes the mass density. The parameter \(a\) controls the relaxation of the polar order parameter \(\mathbf{P}\) and originates from the angular noise of the polar rods. At large \(D_{\mathrm{r}}\), steric interactions between polar rods can be neglected in the dilute regime, and \(a\) is approximately equal to \(D_{\mathrm{r}}\). At small \(D_{\mathrm{r}}\), however, steric-collision--induced alignment cannot be ignored, leading to deviations from this simple relation. In particular, in the limit \(D_{\mathrm{r}} \to 0\), the parameter \(a\) becomes negative due to interaction-induced ordering.
 The term $\lambda \mathbf{v}$ represents polar ordering induced by the flow of the medium. The nonlinear term $-b \lvert \mathbf{P} \rvert^2 \mathbf{P}$ ensures saturation of the order parameter, preventing unbounded growth.
The term $-\Gamma \mathbf{v}$ denotes the dissipative force density arising from friction with the confining plates, while $\alpha \mathbf{P}$ corresponds to the active forcing due to motile rods that generate the velocity field.
Assuming that $\mathbf{v}$ relaxes much faster than $\mathbf{P}$, i.e., $\rho\,\partial_t \mathbf{v} \ll \Gamma\,\mathbf{v},$
we obtain $\mathbf{v} \simeq \frac{\alpha}{\Gamma}\,\mathbf{P}.$ Substituting this expression into Eq.~\eqref{eq:P_eom} yields an effective equation for the polar order parameter,
\begin{equation}
\partial_t \mathbf{P}
\simeq \left( \frac{\alpha\lambda}{\Gamma} - a \right)\mathbf{P}
- b \lvert \mathbf{P} \rvert^2 \mathbf{P}.
\label{eq:effective_P}
\end{equation}
In the steady state, when $\alpha\lambda/\Gamma - a > 0,$
the magnitude of the polar order parameter is given by $\lvert \mathbf{P} \rvert
= \left[ (\alpha\lambda - a\Gamma)/(b\Gamma) \right]^{1/2}.$
For $\alpha\lambda/\Gamma - a < 0,$
the system remains in a disordered state with
$\lvert \mathbf{P} \rvert = 0$. 
The parameters \(a\) and \(\alpha\) depend primarily on \(\phi_{\mathrm{a}}\) and \(D_{\mathrm{r}}\). As the polar rods and the medium rods segregate more strongly, the interfacial length between polar-rod-rich and apolar-rod-rich regions decreases [see Fig.~\ref{fig04}(a)]. Since polar rods push the medium microscopically through frictional contact during collisions, a larger interfacial length allows the polar rods to more effectively push the medium rods. Thus, enhanced segregation reduces $\alpha$, leading to a decrease in the polar order parameter $P$. This provides the physical origin of the anomalous dependence of \(P\) on \(\phi_{\mathrm{a}}\) and \(D_{\mathrm{r}}\). As shown in Fig.~\ref{fig03}(a), the segregation order parameter \(\Sigma\) increases sharply beyond \(\phi_{\mathrm{a}} \simeq 0.6\). As a result, at large \(\phi_{\mathrm{a}}\),  \(P\) decreases with increasing \(\phi_{\mathrm{a}}\) due to the reduced velocity field, in contrast to the behavior observed in a bead medium (see Fig.~\ref{fig05}b).
A similar mechanism operates when varying \(D_{\mathrm{r}}\). At fixed \(\phi_{\mathrm{a}} = 0.50\), \(\Sigma\) decreases with increasing \(D_{\mathrm{r}}\) for small \(D_{\mathrm{r}}\) [see Supplemental Material, Sec.~\ref{sec:SegODPM}, Fig.~\ref{figsegodpmfap50ns}], leading to an increase in \(\alpha\) and a corresponding enhancement of polar ordering [see Fig.~\ref{fig01}(f) and inset of Fig.~\ref{fig05}(b)]. Likewise, \(\Sigma\) increases with the apolar-rod aspect ratio \(\sigma\), resulting in a decrease of \(P\) with increasing \(\sigma\) [see the inset of Fig.~\ref{fig04}(a)], unless \(\sigma\) becomes sufficiently large that the medium itself enters an ordered phase.

The dependence of $a$ on $\phi_{\mathrm{a}}$ becomes particularly important at large $\phi_{\mathrm{a}}$, where nematic currents associated with topological defects enhance the rotational diffusivity of the motile rods (see Supplemental Material, Sec.~\ref{sec:nematic_odpm}, Fig.~\ref{figCtC2t}(c))~\cite{motilerod_AS}. These active currents also disrupt the medium flow generated by the polar rods, further reducing $\alpha$.

\clearpage
\onecolumngrid

\section*{Supplemental Material}

\setcounter{section}{0}
\setcounter{figure}{0}
\renewcommand{\thefigure}{S\arabic{figure}}


\section{Simulation Details}
\label{sec:sim_details}

\begin{figure}[!h]
    \centering
    \includegraphics[width=1\linewidth]{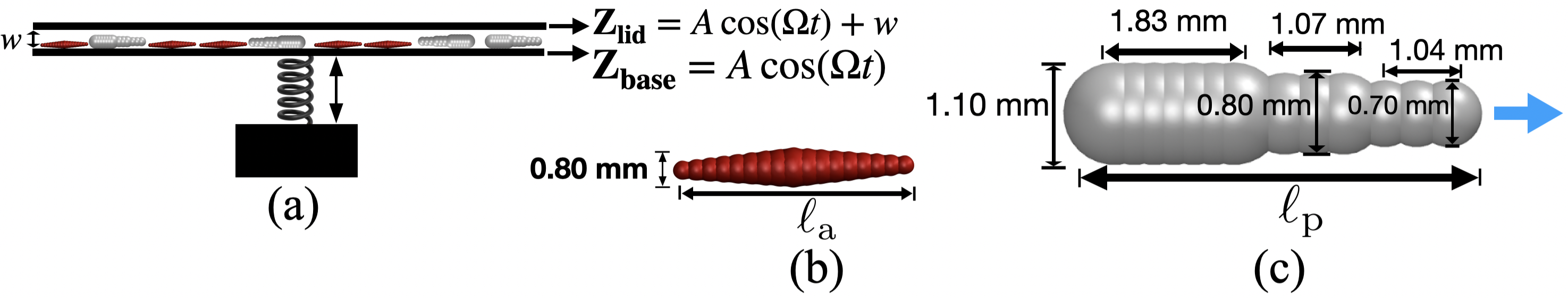}
\caption{(a) Schematic illustration of rods confined between vertically vibrated plates. (b) Geometrical structure of an apolar rod. (c) Geometrical structure of a polar rod, where the blue arrow indicates the direction of self-propulsion.}

    \label{fig01S}
\end{figure}
Although the detailed simulation methodology has been reported previously in Sharma \textit{et al.}~\cite{pahsesAS_HS,motilerod_AS}, we briefly summarize the key aspects here. 
Our system consists of granular rods confined to a quasi-two-dimensional geometry and subjected to vertical sinusoidal vibrations that generate effective horizontal (in-plane) motion (Fig.~\ref{fig01S}a). The system contains geometrically apolar rods (Fig.~\ref{fig01S}b) and geometrically polar rods (Fig.~\ref{fig01S}c).

Collisions between rods, as well as between rods and confining boundaries, are treated as instantaneous events and are implemented using an impulse-based collision model~\cite{kumar2014flocking,2,3,4,5,pahsesAS_HS,motilerod_AS}. The calculation of post-collisional velocities requires the inertia tensors of the particles, expressed in arbitrary mass units. These tensors are evaluated assuming the rods are solid bodies, with the mass ratio of apolar to polar rods fixed at 0.20. 

Between successive collisions, the rods evolve according to Newtonian rigid-body dynamics, following the approach described in~\cite{pahsesAS_HS}. The trajectories are generated using a time-driven particle dynamics algorithm. The friction and restitution coefficients corresponding to the various collision types are provided in Table~\ref{tab:table2}. All simulation snapshots and movies were rendered using the VMD software~\cite{VMD}.

 The area fractions of apolar $\phi_{\mathrm{a}}$ and polar rods $\phi_{\mathrm{p}}$ are defined as:
\begin{equation}
\phi_{\mathrm{a}} = \frac{N_{\mathrm{a}} A}{L^{2}}, \qquad
\phi_{\mathrm{p}}  = \frac{N_{\mathrm{p}} A}{L^{2}},
\end{equation}
where $N_{\mathrm{a}}$ and $N_{\mathrm{p}}$ denote the numbers of apolar and polar rods, respectively, and $A$ is the projected area of a rod lying flat in the $xy$ plane.
The numbers of polar and apolar rods are determined by the chosen area fraction and the length of the apolar rods. Over the parameter range explored, the number of polar rods varies from \( N_{\mathrm{p}} = 801 \) to \( 3203 \), while the number of apolar rods spans according to aspect ratio $\sigma$. The value of $\sigma$ and the range of apolar rods are given in the table~\ref{tab:table1}. 

\begin{table}[hbt!]
\caption{\label{tab:table1}%
Number of apolar rods for varius rod length
}
\begin{ruledtabular}
\begin{tabular}{lcdr}
\textrm{$\sigma$} &
\textrm{   $\ell_{\mathrm{a}}$ (mm)}&
\textrm{$N_{\mathrm{a}}$ range}\\
\colrule

 1.00 (beads) & 0.80  &  6317- 101070\\
1.25 & 1.00  & 5644-84660  \\
1.88 & 1.50   & 3921-39212  \\
2.50 & 2.00  & 2927-29266  \\
3.12 & 2.50   & 2322- 37147 \\
5.62 & 4.50   & 1261-18915 \\
\end{tabular}
\end{ruledtabular}
\end{table}


\begin{table}[hbt!]
\caption{\label{tab:table2}%
Friction ($\mu$) and restitution ($e$) coefficients used for the different collision interactions in the system.
}
\begin{ruledtabular}
\begin{tabular}{lcdr}
\textrm{Collision} &
\textrm{$\mu$} && 
\textrm{$e$} \\
\colrule
Between Rods   & 0.05 && 0.30 \\
Polar-Base/Lid   & 0.03 && 0.10 \\
Apolar-Base/Lid   & 0.01 && 0.30 \\

\end{tabular}
\end{ruledtabular}
\end{table}

To incorporate intrinsic rotational noise in the polar rod, we impose a stochastic angular velocity along the $z$-axis whenever the rod collides with either the base or the lid. The angular velocity is defined as $\omega_z = \varepsilon v_{\text{rel}} \eta,$ where $v_{\text{rel}}$ denotes the normal component of the relative velocity at the contact point, $\varepsilon$ is a control parameter that sets the noise strength, and $\eta$ is a random variable taking values $\pm 1$ with equal probability. This procedure effectively generates rotational diffusion in the rod’s orientation, characterized by a rotational diffusion coefficient $D_r$, which increases with $\varepsilon$. Throughout this work, $D_r$ is used to quantify the stochastic fluctuations in the orientation of the polar rod.

\section{Order Parameters and Correlations}
\label{sec:nematic_odpm}

\textbf{Polar and nematic order parameters:}
The polar and nematic ordering is quantified using the polar ($P$) and nematic ($S$) order parameters, respectively defined as:
\begin{equation}
P=\sqrt{\left\langle\cos \theta_{\text{p}_i}\right\rangle^2+\left\langle\sin \theta_{\text{p}_i}\right\rangle^2}; S=\sqrt{\left\langle\cos 2\theta_{\text{a}_i}\right\rangle^2+\left\langle\sin 2\theta_{\text{a}_i}\right\rangle^2}
\end{equation} 
Here, $\theta_{\mathrm{p}_i}$ and $\theta_{\mathrm{a}_i}$ denote the orientations of the $i$th polar and apolar rods, respectively, measured relative to a fixed reference axis. The notation $\langle \cdot \rangle$ represents the ensemble average taken over the corresponding set of rods.
\begin{figure}[!h]
    \centering
     \includegraphics[width=0.9\textwidth]{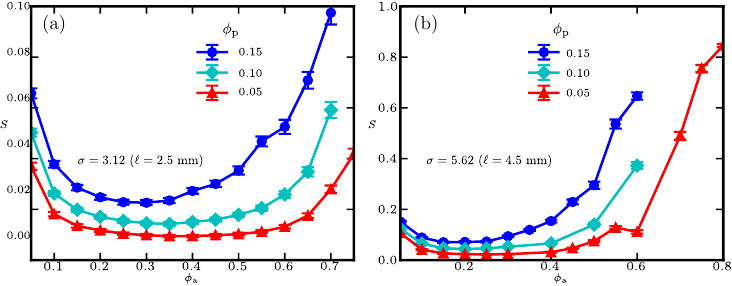}
    \caption{(a) $S$ vs $\phi_{\mathrm{a}}$ for $\sigma=3.12$  and (b) $S$ vs $\phi_{\mathrm{a}}$ for $\sigma=5.62$. In the absence of intrinsic angular noise in the polar rods, i.e., $D_{\mathrm{r}} = 0$ rad$^2$ s$^{-1}$.}
    \label{figS}
\end{figure}

Fig.~\ref{figS}(a) presents the variation of the nematic order parameter $S$ with $\phi_{\mathrm{a}}$ at different  with $\phi_{\mathrm{p}}$  for $\sigma = 3.12$ ($\ell_{\mathrm{a}} = 2.5$ mm) and $D_{\mathrm{r}} = 0$ rad$^2$ s$^{-1}$. 
Although the apolar medium remains in an isotropic phase over the entire range of $\phi_{\mathrm{a}}$, the nematic order parameter of the system exhibits a non-monotonic behavior. Specifically, $S$ initially decreases with increasing $\phi_{\mathrm{a}}$, and subsequently increases at larger values of $\phi_{\mathrm{a}}$. At low $\phi_{\mathrm{a}}$, the density of apolar rods is small, resulting in relatively weak interactions between the polar rods and the surrounding medium. As $\phi_{\mathrm{a}}$ increases, frequent collisions with the medium rods disrupt the orientational coherence of the polar rods, leading to a reduction in $S$. However, at higher $\phi_{\mathrm{a}}$, enhanced interactions promote local nematic alignment among the polar rods, which in turn increases the overall nematic order parameter. Throughout this range, the apolar medium itself does not develop global nematic order and remains isotropic.

Fig.~\ref{figS}(b) displays the variation of $S$ as a function of $\phi_{\mathrm{a}}$ at different $\phi_{\mathrm{p}}$ for $\sigma = 5.62$ ($\ell_{\mathrm{a}} = 4.5$ mm), and $D_{\mathrm{r}} = 0$ rad$^2$ s$^{-1}$. 
We observe a qualitatively similar non-monotonic behavior to that seen for $\sigma = 3.12$. However, in this case, as $\phi_{\mathrm{a}}$ increases, the apolar medium undergoes a transition from the isotropic to the nematic phase.

\textbf{Polar and nematic order parameter correlation function:}
To investigate the existence of long-range order, we evaluate the polar and nematic orientational correlation functions, $G(r)$ and $G_2(r)$, corresponding to the polar and apolar rods, respectively, defined as:
\begin{equation}
G(r) = \left\langle\cos\left[(\theta_{\text{p}_i} - \theta_{\text{p}_j})\right]\right\rangle_r; G_2(r) = \left\langle\cos\left[2(\theta_{\text{a}_i} - \theta_{\text{a}_j})\right]\right\rangle_r
\end{equation}
where $\langle \cdot \rangle_r$ denotes an average over all pairs of rods of the same type separated by a distance $r$. Here, $\theta_{\text{p}_i}$ and $\theta_{\text{p}_j}$ are the orientations of the $i$th and $j$th polar rods, and $\theta_{\text{a}_i}$ and $\theta_{\text{a}_j}$ correspond to those of the apolar rods, measured with respect to a fixed reference axis.
\begin{figure}[!h]
    \centering
     \includegraphics[width=0.9\textwidth]{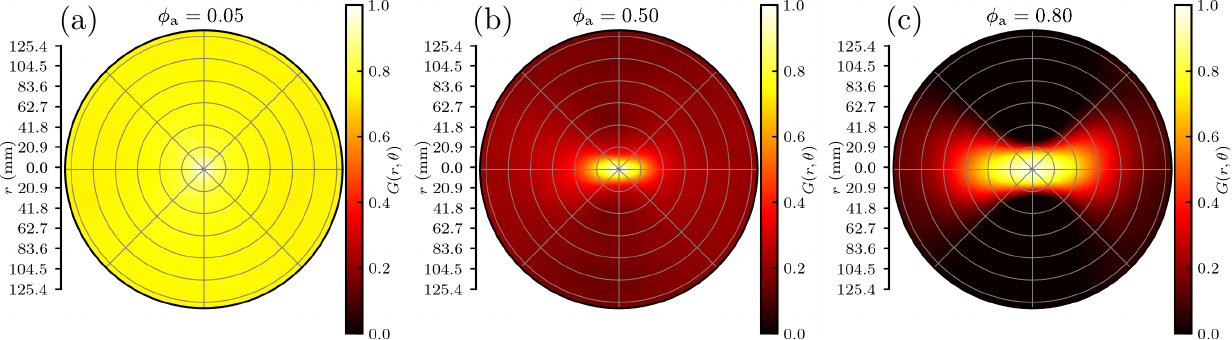}
    \caption{$G(r,\theta)$ (a) for $\phi_{\mathrm{a}}=0.05$ (b) $\phi_{\mathrm{a}}=0.50$ and (c) $\phi_{\mathrm{a}}=0.80$. For $\sigma = 3.125$ ($\ell_{\mathrm{a}} = 2.5,\mathrm{mm}$), $D_{\mathrm{r}} = 0~\mathrm{rad}^2,\mathrm{s}^{-1}$, and fixed $\phi_{\mathrm{p}}=0.05$ .}
    \label{figGrtheta}
\end{figure}

To examine the anisotropy of the polar order parameter correlation function of polar rods, we compute $G(r,\theta)$ in the comoving frame of a reference polar rod and average over all polar rods. Fig.~\ref{figGrtheta} present polar heat maps of $G(r,\theta)$ for $\sigma = 3.12$ ($\ell_{\mathrm{a}} = 2.5~\mathrm{mm}$), $\phi_{\mathrm{p}} = 0.05$, and $D_{\mathrm{r}} = 0~\mathrm{rad}^2\,\mathrm{s}^{-1}$. In Fig.~\ref{figGrtheta}(a), corresponding to $\phi_{\mathrm{a}} = 0.05$, $G(r,\theta)$ is nearly isotropic, indicating circular symmetry around the reference rod. In Fig.~\ref{figGrtheta}(b), for $\phi_{\mathrm{a}} = 0.50$, the correlation function becomes anisotropic. In Fig.~\ref{figGrtheta}(c), at $\phi_{\mathrm{a}}=0.80$, the anisotropy becomes more pronounced. This behavior arises because, at high $\phi_{\mathrm{a}}$, the flocks become longitudinally extended. As a result, $G(r,\theta)$ exhibits strong anisotropy, with enhanced correlations of polar rods along the direction of motion (ahead of and behind the reference rod) over long distances.

\textbf{Polar and nematic order parameter autocorrelation function:}
The statistical independence of configurations is evaluated through the orientational autocorrelation functions $C(t)$ and $C_2(t)$ for the polar and apolar rods, respectively, defined as:
\begin{equation}
C(t) = \left\langle\cos\left\{(\theta_{\text{p}_i}(t+\tau) - \theta_{\text{p}_i}(t))\right\}\right\rangle_i; C_2(t) = \left\langle\cos\left\{2(\theta_{\text{a}_i}(t+\tau) - \theta_{\text{a}_i}(t))\right\}\right\rangle_i 
\end{equation}
where the average $\langle \cdot \rangle_i$ is taken over all rods of the corresponding type. Here, $\theta_{\text{p}i}(t)$ and $\theta_{\text{p}i}(t+\tau)$ represent the orientations of the same $i^{\text{th}}$ polar rod at times $t$ and $(t+\tau)$, respectively, and $\theta_{\text{a}i}(t)$ and $\theta_{\text{a}_i}(t+\tau)$ correspond to those of the apolar rod.
\begin{figure}[!h]
    \centering
     \includegraphics[width=0.9\textwidth]{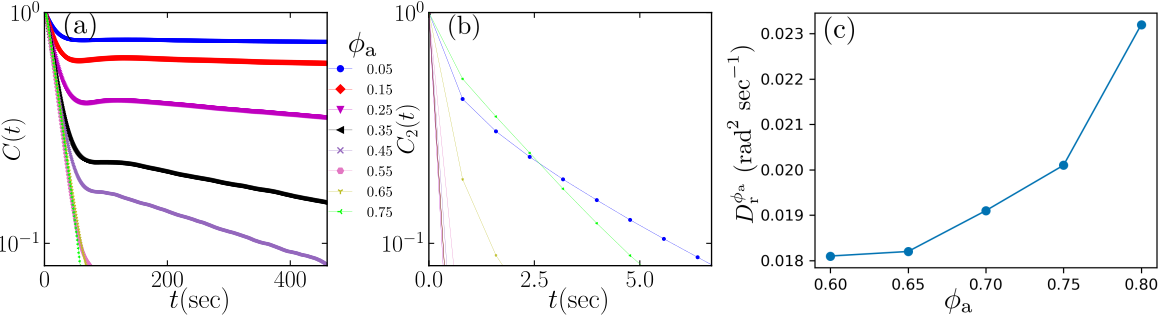}
    \caption{(a) Semilog plot of polar rod orientation autocorrelation $C(t)$ for different $\phi_\text{a}$ and (b) apolar rod orientation autocorrelation $C_2(t)$ for different $\phi_\text{a}$ and (c) Rotational diffusion of polar rods $D_{\text{r}}^{\phi_{\text{a}}}$, correspoing to C(t).The parameter for the above is $\ell_{\text{a}}=2.5$ mm, $\phi_{\text{p}}=0.05$ and $D_{\text{r}}=0 $ rad$^2$ sec$^{-1}$.}
    \label{figCtC2t}
\end{figure}
Fig.~\ref{figCtC2t} presents the autocorrelation function of polar rod and medium rods for medium rod length ($\ell_\text{a}=2.5$ mm) at fixed $\phi_\text{p}=0.05$ and $D_{\text{r}}=0 $ rad$^2$ sec$^{-1}$. In Fig.~\ref{figCtC2t}(a), $C(t)$ saturates to a nonzero value for small $\phi_\text{a}$ and decays for $\phi_\text{a} > 0.55$, reflecting a transition from coherent motion of polar rods to randomly moving flocks that continuously break and reform. Rotational diffusion for polar rods is shown Fig.~\ref{figCtC2t}(c) it increases with $\phi_{\text{a}}$. Fig.~\ref{figCtC2t}(b) shows the autocorrelation function of medium rods $C_2(t)$, which displays a rapid decay with time for all $\phi_\text{a}$.

\section{Cluster Analysis}
\label{sec:clusterana}
\begin{figure}[!h]
    \centering
    \includegraphics[width=\linewidth]{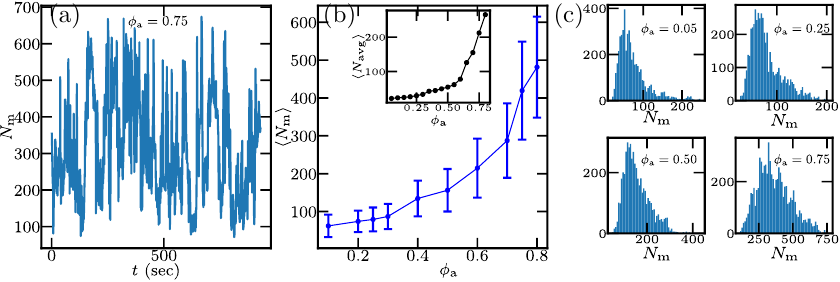}
    \caption{ (a) The largest flock size $N_{\mathrm{m}}$ vs $t$ at $\phi_{\mathrm{a}} = 0.75$. (b) Steady-state averaged largest flock size $\langle N_{\mathrm{m}} \rangle$ vs $\phi_{\mathrm{a}}$; Inset: average flock size $N_{\mathrm{avg}}$ vs $\phi_{\mathrm{a}}$. (c) Histogram of  $N_{\mathrm{m}}$ for different $\phi_{\mathrm{a}}$. Here $\phi_{\text{p}}=0.05$ and $D_{\text{r}}$=0 rad$^2$ sec$^{-1}$.}
    \label{figlg}
\end{figure}

To quantify collective clustering, we employ the unsupervised machine learning algorithm DBSCAN (Density-Based Spatial Clustering of Applications with Noise) to identify flocks and determine the largest cluster size. Using this method, we extract the largest polar flock in an isotropic medium at $\sigma = 3.125$ ($\ell_{\mathrm{a}} = 2.5,\mathrm{mm}$), in the absence of intrinsic angular noise ($D_{\mathrm{r}} = 0~\mathrm{rad}^2,\mathrm{s}^{-1}$) and at fixed $\phi_{\mathrm{p}} = 0.05$, as shown in Fig.~\ref{figlg}. Figure~\ref{figlg}(a) displays the time evolution of the largest flock size $N_{\mathrm{m}}$ for $\phi_{\mathrm{a}} = 0.75$. The pronounced temporal fluctuations reflect the highly dynamic nature of the flocks, which continuously merge and fragment. Figure~\ref{figlg}(b) shows the steady-state averaged largest cluster size $\langle N_{\mathrm{m}} \rangle$ as a function of $\phi_{\mathrm{a}}$, with error bars denoting temporal fluctuations. The inset presents the average flock size $N_{\mathrm{avg}}$, obtained from the closed contour corresponding to $70\%$ of the maximum value of $g(r,\theta)$ [black contours in Figs. 4(b) and (c) of the main text]. Both $\langle N_{\mathrm{m}} \rangle$ and $N_{\mathrm{avg}}$ increase monotonically with $\phi_{\mathrm{a}}$, indicating enhanced clustering at higher medium densities.
Fig.~\ref{figlg}(c) shows the distribution of flock sizes for different $\phi_{\mathrm{a}}$, exhibiting a systematic shift toward larger $N_{\mathrm{m}}$ with increasing $\phi_{\mathrm{a}}$.

\section{Segragation Order Parameter}
\label{sec:SegODPM}
In order to quantify  how well-mixed are polar rods inside medium we compute segragation order parameter $\Sigma$ defined 
as~\cite{mccandlish2012spontaneous,bhowmik2025segregation}:
\begin{equation}
\Sigma =
\frac{1}{2\,\phi_{\mathrm{p}}\,[1-\phi_{\mathrm{p}}]}
\sum_{i=1}^{N}
\left(
\frac{n_i}{n_{\mathrm{tot}}}
\right)
\left|
\phi_i - \phi_{\mathrm{p}}
\right|,
\label{eq:order_parameter}
\end{equation}
where simulation box is devided into $N$ number of square cells of side length $a$. where $\phi_i$ is fraction of polar rods in the $i$th cell, $n_i$ is the total number of polar rods in the $i$th cell, $n_{\mathrm{tot}}$ is the total number of polar rods in the simulation box, and $\phi_{\mathrm{p}}$ is the total fraction of the polar rods in the simulation box. The prefactor normalizes $\Sigma$ such that $\Sigma \in [0,1]$, where $\Sigma = 0$ corresponds to the theoretically minimal level of segregation, while $\Sigma = 1$ indicates complete (maximal) segregation.
Noting that $\Sigma$ is nonzero even in the absence of the medium, we subtract this baseline contribution in order to compute the value of $\Sigma$ in the presence of the medium.
\begin{figure}
    \centering
    \includegraphics[width=0.3\linewidth]{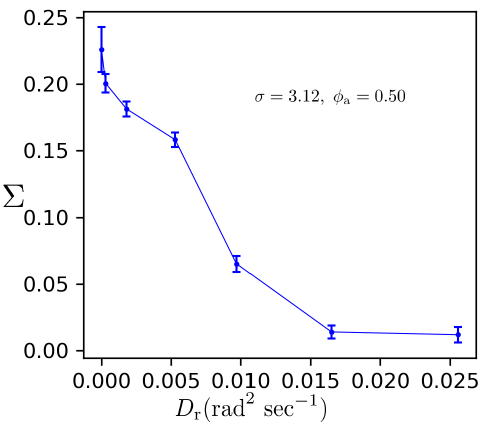}
    \caption{$\Sigma$ vs $D_{\mathrm{r}}$. Here $\sigma=3.125$, $\phi_{\text{a}}=0.50$ and $\phi_{\text{p}}=0.05$. }
    \label{figsegodpmfap50ns}
\end{figure}
\section{Polar rods in spherical bead medium}
\label{sec:polar_sphericalbeads}
\begin{figure}[!h]
    \centering
    \includegraphics[width=\linewidth]{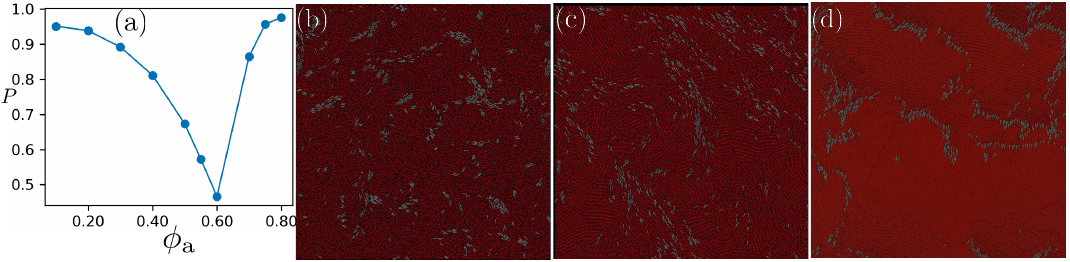}
    \caption{(a) $P$ vs $\phi_{\mathrm{a}}$ for $\sigma=1.0$ (the bead medium).  Steady-state configurations: (b) $\phi_{\mathrm{a}}=0.60$, (c) $\phi_{\mathrm{a}}=0.70$ and (d) $\phi_{\mathrm{a}}=0.75$. Here, $\phi_{\mathrm{p}}=0.05$  and $D_{\mathrm{r}} = 0~\mathrm{rad}^2\mathrm{s}^{-1}$.}
    \label{figbead}
\end{figure}

Fig.~\ref{figbead}(a) presents the variation of $P$ with $\phi_{\mathrm{a}}$ for a bead medium of diameter $0.80\,\mathrm{mm}$ at fixed $\phi_{\mathrm{p}}=0.05$ and $D_{\mathrm{r}} = 0~\mathrm{rad}^2\,\mathrm{s}^{-1}$. The order parameter exhibits a non-monotonic dependence on $\phi_{\mathrm{a}}$: it decreases initially, attains a minimum at $\phi_{\mathrm{a}}=0.60$, and subsequently increases at higher area fractions. The corresponding steady-state configurations are shown in Figs.~\ref{figbead}(b)–(d). At $\phi_{\mathrm{a}}=0.60$ [Fig.~\ref{figbead}(b)], the system is disordered, with no global alignment among the polar rods. Increasing the area fraction to $\phi_{\mathrm{a}}=0.70$ [Fig.~\ref{figbead}(c)] restores collective motion, characterized by coherent alignment and a spatially homogeneous distribution. At $\phi_{\mathrm{a}}=0.75$ [Fig.~\ref{figbead}(d)], the rods self-organize into finite-sized bands that propagate coherently through the medium.

\section{Polar rods in nematic phase}
\label{sec:innematicphase}
\begin{figure}
    \centering
    \includegraphics[width=\linewidth]{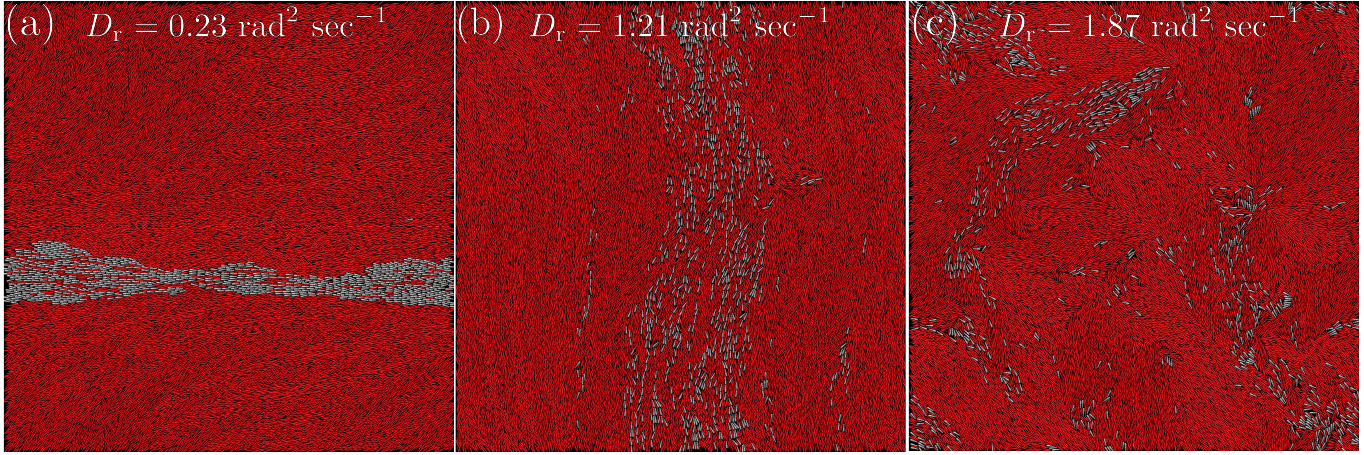}
    \caption{Steady-state configurations in the nematic phase for (a) $D_{\mathrm{r}} = 0.23~\mathrm{rad}^2\,\mathrm{s}^{-1}$, 
(b) $D_{\mathrm{r}} = 1.21~\mathrm{rad}^2\,\mathrm{s}^{-1}$, and 
(c) $D_{\mathrm{r}} = 1.87~\mathrm{rad}^2\,\mathrm{s}^{-1}$. 
Parameters: $\sigma = 5.62$, $\phi_{\mathrm{a}} = 0.75$, and $\phi_{\mathrm{p}} = 0.05$.}
    \label{figl45n15n30n36}
\end{figure}

When the medium is in the nematic phase at high area fraction ($\sigma = 5.62$), the polar rods organize into longitudinal flocks aligned with the nematic director. Upon introducing rotational noise $D_{\mathrm{r}}$, these flocks develop density modulations along the director and progressively become dilute. With further increase in $D_{\mathrm{r}}$, the bands lose coherence and transition to a disordered state. Simultaneously, the surrounding medium also becomes disordered due to the generation of topological defects induced by the noise. For Fig.~\ref{figl45n15n30n36}, $\sigma = 5.62$, $\phi_{\mathrm{a}} = 0.75$, and $\phi_{\mathrm{p}} = 0.05$. 
Fig.~\ref{figl45n15n30n36}(a) shows densely packed bands at $D_{\mathrm{r}} = 0.23~\mathrm{rad}^2\,\mathrm{s}^{-1}$. 
In Fig.~\ref{figl45n15n30n36}(b), the structure becomes dilute at $D_{\mathrm{r}} = 1.21~\mathrm{rad}^2\,\mathrm{s}^{-1}$. 
At $D_{\mathrm{r}} = 1.87~\mathrm{rad}^2\,\mathrm{s}^{-1}$ (Fig.~\ref{figl45n15n30n36}(c)), the bands lose their integrity and the medium exhibits strong disorder accompanied by topological defects.
\section{DESCRIPTION OF SUPPLEMENTARY MOVIES}
\noindent \textbf{Simulation Movies:} 
\textbf{\href{https://drive.google.com/drive/folders/1LGRRIl93ZgJlMpLSHNNaBJ-fJL3bOjlF?usp=share_link}{Url for simulation movies}}\\
\noindent \textbf{File name SM1:} Simulation Movie For $\sigma = 3.125$ ($\ell_{\mathrm{a}} = 2.5,\mathrm{mm}$), $D_{\mathrm{r}} = 0~\mathrm{rad}^2,\mathrm{s}^{-1}$, and fixed $\phi_{\mathrm{p}}=0.05$. \textbf{Left:} $\phi_{\mathrm{a}}=0.05$, \textbf{Observation:} Polar rods move coherently by making small swarms, \textbf{Right:} $\phi_{\mathrm{a}}=0.80$, \textbf{Observation:} Polar rods form longitudinal flocks, continuously breaking into smaller flocks that subsequently merge to form larger polar flocks.

\noindent \textbf{File name SM2:} Simulation movie for $\sigma = 3.125$ 
($\ell_{\mathrm{a}} = 2.5\,\mathrm{mm}$), 
$\phi_{\mathrm{a}} = 0.50$, and fixed $\phi_{\mathrm{p}} = 0.05$. 
\textbf{Left:} $D_{\mathrm{r}} = 0\,\mathrm{rad}^2\,\mathrm{s}^{-1}$, \textbf{Observation:} Randomly moving, large-sized longitudinal flocks reduce the global polar order parameter. \textbf{Right:} $D_{\mathrm{r}} = 0.026\,\mathrm{rad}^2\,\mathrm{s}^{-1}$, \textbf{Observation:} Large size longitudinal flock breaks into smaller flocks and moves coherently, enhancing the global polar order parameter. 

\noindent \textbf{File name SM3:} Simulation movie for different $\sigma$ values at fixed $\phi_{\mathrm{a}} = 0.50$, $\phi_{\mathrm{p}} = 0.05$ and $D_{\mathrm{r}} = 0.23\,\mathrm{rad}^2\,\mathrm{s}^{-1}$.\\
\textbf{Top Left:} $\sigma=1.00$ (spherical bead medium), \textbf{observation:} Polar rods form transverse folck.
\textbf{Top Right:} $\sigma=1.25$ (slighly apolar rod medium), \textbf{observation:} Polar rods moves homogeneously in the medium.
\textbf{Bottom Left:} $\sigma=3.12$, \textbf{observation:} Polar rods form randomly moving flock of smaller size. 
\textbf{Bottom Right:} $\sigma=5.63$, \textbf{observation:} Polar rods form a single longitudinal flock which moves along the nematic director. 

\noindent \textbf{File name SM4:} Simulation movie for $\sigma=5.62$ ($\ell_{\mathrm{a}}=4.5$ mm) values at fixed $\phi_{\mathrm{a}} = 0.75$, $\phi_{\mathrm{p}} = 0.05$ and $D_{\mathrm{r}} = 0.23\,\mathrm{rad}^2\,\mathrm{s}^{-1}$
(medium in nematic phase). \textbf{observation:} Small flock of polar rods merge into large flock due to emergence of +1/2 defect.

\noindent \textbf{File name SM5:} Simulation movie for $\sigma=3.12$ ($\ell_{\mathrm{a}}=2.5$ mm) values at fixed $\phi_{\mathrm{a}} = 0.70$, $\phi_{\mathrm{p}} = 0.05$ and $D_{\mathrm{r}} = 0 \,\mathrm{rad}^2\,\mathrm{s}^{-1}$ \textbf{observation:} Polar rods shows milling like behaviour.


\end{document}